\pdfoutput=1
\documentclass[sigconf, nonacm]{acmart}

\usepackage{pvldb}%

\renewcommand\vldbdoi{XX.XX/XXX.XX}
\renewcommand\vldbpages{XXX-XXX}
\renewcommand\vldbavailabilityurl{} %

\usepackage{amsmath}

\usepackage{amssymb}
\usepackage{booktabs}
\usepackage{multirow}
\usepackage{xcolor}
\usepackage{enumitem}
\usepackage{microtype}
\usepackage{graphicx}
\usepackage{xspace}
\usepackage{needspace}
\usepackage{tikz}
\usepackage{algorithm}
\usepackage{algpseudocode}
\usepackage[normalem]{ulem}
\usepackage{xcolor}

\usetikzlibrary{arrows.meta,positioning}
\definecolor{RRBlue}{HTML}{1D4ED8}
\definecolor{RROrange}{HTML}{C2410C}
\definecolor{RRGreen}{HTML}{047857}
\definecolor{RRGray}{HTML}{4B5563}

\newcommand{\system}{\textsc{ARoute}\xspace}
\newcommand{\systemB}{\textsc{ARoute}-Batch\xspace}
\newcommand{\systemS}{\textsc{ARoute}-Stream\xspace}

\ifdefined\cleanbuild
 \newcommand{\del}[1]{}
 
\else
 \newcommand{\del}[1]{{\color{gray}\sout{#1}}}
 
\fi

\begin{document}

\title{ARoute: Task-Aligned Batch and Streaming Model Routing}

\author{Daofu Zhang}
\affiliation{%
 \institution{University of Utah}
 \city{Salt Lake City}
 \country{United States}
}
\email{daofu.zhang@utah.edu}

\author{Qingcheng Zeng}
\affiliation{%
 \institution{Northwestern University}
 \city{Evanston}
 \country{United States}
}
\email{qingchengzeng2027@u.northwestern.edu}
\author{Chunwei	Liu}
\affiliation{%
 \institution{Purdue University}
 \city{West Lafayette}
 \country{United States}
}
\email{	chunwei@purdue.edu}
\author{Fuheng	Zhao}
\affiliation{%
 \institution{University of Utah}
 \city{Salt Lake City}
 \country{United States}
}
\email{fuheng.zhao@utah.edu}

\begin{abstract}
Language models increasingly power data-intensive applications, from semantic data processing to question answering and agentic workflows. Using frontier models for every request is prohibitively expensive, while relying only on smaller models reduces accuracy. We study model routing under a user-specified cost budget, aiming to maximize accuracy under task-specific correctness criteria. Existing routers fall short in two ways: pretrained routers rely on generic preference signals that may not reflect a task's own correctness criteria or a new model pair, and model cascades run the weaker model first on every request, adding sequential latency. We present \system, a task-aligned router that learns cost and accuracy signals to select a model for each request before inference. \system has two variants: \systemB operates on a batch of requests known in advance, routing requests with the greatest expected benefit to the stronger model subject to the budget, whereas \systemS processes requests online, using a threshold calibrated on a small validation set to satisfy the cost quota and dynamically adjusting it as the budget is consumed. For \systemS, we provide a finite-sample guarantee on expected normalized spending. We evaluate \system on MMLU-Pro, LiveCodeBench, SuperGPQA, and two value-filling tasks built from SWAN semantic query benchmark against other state-of-the-art approaches. At matched cost, \systemB achieves the highest mean accuracy in all 15 task--cost settings, improving over the strongest baseline by 4.1 percentage points on average and by up to 11.9 points. \system also completes a batch up to 2.8$\times$ faster than cascades.
\end{abstract}

\maketitle

\pagestyle{\vldbpagestyle}

\section{Introduction}
\label{sec:intro}
{\color{black}
Language models increasingly serve as execution components in data
systems and agent workloads. In semantic data processing, semantic operators such as semantic filters and semantic joins use large language models (LLMs) to process unstructured data \cite{patel2025lotus,liu2025palimpzest,dai2024uqe,shankar2025docetl,arora2023evaporate,anderson2025aryn,lin2025zendb,biswal2025tag,lao2026sembench}.
In agent workloads, sequences of LLM requests are issued to complete multi-step reasoning tasks \cite{yao2023react,khattab2024dspy}. Both settings can require large
numbers of model calls, making the choice of model for each one an important determinant of cost and accuracy.
Using frontier language models for every request can make these workloads prohibitively expensive \cite{huang2025thriftllm,jo2025sparellm}, whereas relying entirely on smaller models can sacrifice accuracy. Model routing addresses this tradeoff by selecting a model for each request, directing calls to stronger models when the expected accuracy gains justify the additional cost \cite{hu2024routerbench,varangot2025survey,ong2025routellm,chen2024frugalgpt}.
In this work, we study model routing with the goal of maximizing accuracy within a given cost budget.\par}

\begin{figure}[t]
\centering
\includegraphics[width=\columnwidth]{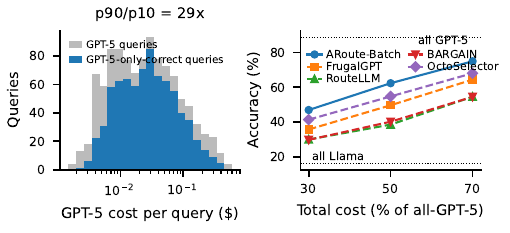}
\caption{LiveCodeBench. Left: GPT-5 cost per query (p90/p10 = 29$\times$); queries only GPT-5 answers correctly (OSC) occur at every price. Right: accuracy at matched total cost, paired training records charged.}
\label{fig:teaser}
\end{figure}

{\color{black}
The central question is which requests benefit from the stronger model
and which the weaker model already answers correctly.
Figure~\ref{fig:teaser} (right) shows the stakes on LiveCodeBench \cite{jain2025livecodebench}:
answering every query with Llama \cite{grattafiori2024llama3} gives 16\% accuracy and answering
every query with GPT-5 \cite{openai2025gpt5} gives 89\%, and a router recovers part of that
gap at a fraction of the all-GPT-5 cost, but how much depends on which
queries it upgrades. Consider a semantic filter that identifies
customer reviews reporting product defects. A weaker model classifies
an explicit statement such as ``the screen arrived cracked'' correctly,
but it misses a defect described only indirectly (``worked for two
days, then nothing''), which the stronger model catches; only the
second review is worth an upgrade. The price of an upgrade varies as
well. Figure~\ref{fig:teaser} (left) plots GPT-5's cost per query on
the same task: it spans a 29$\times$ range between the 10th and 90th
percentiles because output lengths differ \cite{zheng2023lengthperception}, and the queries that only
GPT-5 answers correctly occur at every price. When two requests promise
similar accuracy gains, upgrading the cheaper one leaves more of the
budget for other requests. Effective routing therefore requires
estimating both the accuracy benefit and the additional cost of the
stronger model for each request.\par}

{\color{black}
Both estimates must be aligned with the task. An upgrade pays off only
when the weaker model answers wrong and the stronger one answers right,
and whether an answer counts as right is decided by the task's own
criterion. Which queries meet this condition also depends on the two
models in use: swap in a different weaker model and the set changes.
The price of an upgrade is task-specific as well: under token pricing
it is set by how many tokens each model generates, and a request for a
program yields far more output than a request for a label. Training data prepared in advance therefore does not fit: a
routing log collected on other tasks or other model pairs answers a
different question, and the only evidence that fits comes from the
user's own workload. A sample of requests is answered by both models,
each answer is judged under the task's criterion, and the two calls
record what each model cost. Of the four possible outcomes of such a
paired record, only one, weak wrong and strong right, marks a query
worth upgrading.

Producing these records is the cost of alignment, and
it should be as small as possible. A real workload usually has no free
oracle: benchmarks judge answers through reference values or tests, but
a semantic filter encodes a predicate the user wrote and a value-filling
task asks for fields only the user can verify, so the user must read
the query and both answers and decide. Each record also pays two model
calls, one of them to the stronger model, out of the budget the router
is meant to save. We therefore design \system to align with a task from
a small set of such records. From these few labels the router must
predict, for every remaining request and before either model is
invoked, whether the stronger model will help and how much more it will
cost. Each prediction poses its own challenge.}

{\color{black}\textbf{First, the only-strong-correct outcome must be learned for this task and model pair.} For a fixed model pair the useful signal is not which model
is better overall but on which queries the weaker model fails and the
stronger one succeeds; we call such queries \emph{only-strong-correct}
(OSC). OSC queries are common, but a router pretrained on general
preference data \cite{ong2025routellm,ding2024hybridllm} does not find them, nor does the weak model's own confidence \cite{gupta2024cascades,jitkrittum2023confidence}. On our five tasks, RouteLLM's
score ranks a random OSC query above a random non-OSC query only
45--61\% of the time, against 50\% for a random ordering
(Section~\ref{sec:main-results}). Without this signal a router pays for
upgrades that change no answer.}

{\color{black}\textbf{Second, the price of each upgrade must be predicted before
generation.} A budget buys upgrades, so what matters is the benefit per
dollar: between two queries with the same probability of benefit, the
one whose upgrade costs a tenth as much buys ten times the expected
gain, and the dollars it saves pay for further upgrades. The output
length that sets a call's cost is known only after the model has
answered \cite{zheng2023lengthperception,jin2023s3,fu2024learningtorank}, so the incremental cost of the stronger model must be
estimated from the query alone, and that cost varies widely, as the
29$\times$ spread of Figure~\ref{fig:teaser} (left) shows. A router
without this estimate can exhaust its budget on a few expensive
upgrades that correct a handful of answers while cheaper upgrades would
have corrected many more.}

{\color{black}
\textbf{Prior work and limitations.}
No existing router meets both challenges at the query level from a
few labels. Cascades such as FrugalGPT and BARGAIN read the weaker
model's answer before deciding whether to escalate
\cite{chen2024frugalgpt,gupta2024cascades,zeighami2025bargain,aggarwal2024automix,yue2024mot,nie2024onlinecascade}; that
answer is task-specific evidence, but it is bought with a weak call and
a wait on every query, and neither system predicts what the stronger
call will cost. Pre-generation routers (routers that pick the model
from the query alone, before any model generates an answer)
\cite{lu2024zooter,chen2024routerdc,feng2025graphrouter,sakota2024forc,somerstep2025carrot} such as
RouteLLM avoid the extra call, but their released score is trained on
general preference data and treats every call as one unit of cost
\cite{ong2025routellm}. OctoSelector fits task-specific estimates, but
per cluster of queries, so it cannot tell the queries in a cluster
apart \cite{zhang2026octoselector}.
What is missing is a router
that, like RouteLLM, decides before generation, and, like a cascade,
uses task-specific evidence, but obtains that evidence from a small set
of paired records instead of a weak call on every query or thousands of labeled queries \cite{ding2024hybridllm,chen2024routerdc,zhang2026octoselector}, and scores each query on its own rather than by cluster.\par}

{\color{black}
\textbf{Our approach.} We propose \system, a pre-generation router that
learns both signals from the same small set of paired records. Our key
observation is that the two challenges share one input: whether the
stronger model will help and how much more it will cost are both
functions of the query text, so one representation can serve both
predictions. Specifically, \system embeds each query once with a frozen
BERT encoder \cite{devlin2019bert} and trains two lightweight heads on that embedding: a
linear head that predicts the OSC probability (the first challenge) and
a small network that predicts each model's generation cost
(the second challenge). Queries are ranked by predicted OSC probability
per predicted incremental dollar, and those at the top are sent to the
stronger model. In contrast to cascades, \system never runs the weaker
model to decide, so an escalated query costs one call and incurs no
sequential delay. In contrast to pretrained routers, its two heads are
fit to the task's own correctness criterion and prices, and because the
encoder is frozen this takes seconds on a CPU from a small set of records. The same
ranking serves two modes: \systemB ranks a known batch and places its budget cutoff after the fact, from realized costs,
and \systemS certifies a cutoff from a small validation sample with a
finite-sample normalized-cost guarantee, then re-selects it heuristically as the quota is spent. Section~\ref{sec:method} gives the
predictors, the score, and the operating points.\par}
{\color{black}
\textbf{Contributions.} To summarize, our contributions are as follows.
\begin{itemize}[leftmargin=*,topsep=3pt,itemsep=2pt,parsep=0pt]
\item We present \system, a pre-generation router that fits the
 only-strong-correct probability and the incremental cost of each query
 to the task and model pair at hand from a small set of paired records,
 and routes by their ratio before either model is called
 (Sections~\ref{sec:feedback} and~\ref{sec:method}).
\item We show that a frozen encoder with two lightweight heads suffices:
 adapting to a new task takes seconds on a CPU, and once the cost of
 acquiring the records is charged against the budget, a small sample is
 not a compromise but the best choice (Section~\ref{sec:scaling}).
\item We give a finite-sample guarantee on
 the spending of a pre-generation router: a bound on the probability
 that \systemS's validation-selected cutoff exceeds an expected
 normalized-cost quota, from paired costs alone
 (Section~\ref{sec:streaming-case}); prior guarantees concern
 accuracy or a cascade's per-query cost
 \cite{zeighami2025bargain,uddin2026conformalrouting,valkanas2025c3po}.
\item On MMLU-Pro, LiveCodeBench, SuperGPQA, and two SWAN value-filling
 tasks \cite{wang2024mmlupro,jain2025livecodebench,du2025supergpqa,zhao2025swan},
 against RouteLLM, FrugalGPT, BARGAIN, and OctoSelector with all
 acquisition and cascade calls charged, \systemB attains the highest
 mean accuracy in all 15 task--cost settings, exceeding the strongest
 baseline by 4.1 points on average and up to 11.9, and completes a
 batch up to 2.8$\times$ faster than cascades
 (Section~\ref{sec:experiments}).
\end{itemize}
}
\section{Background and Problem Setting}
\label{sec:overview}
In the semantic operators and agent workloads of Section~\ref{sec:intro},
every record or step costs one model invocation, and which model answers
it drives both cost and accuracy. This section makes that per-invocation
choice precise. We call the complete input to one invocation a
\emph{query} $x$: for example, one review together with the predicate
it is tested against, one database record and the field to be filled,
or one programming problem. Each query's correctness and generation
cost are scored in isolation and queries are weighted equally; how an
error propagates through an operator plan or an agent trajectory is
outside our scope. Section~\ref{sec:problem} states the routing
problem, and Sections~\ref{sec:existing} to~\ref{sec:pipeline-overview}
position existing policies against it, state how \system is aligned
with a task, and give a roadmap.

\subsection{Problem formulation}
\label{sec:problem}
Consider a fixed weak model $M_w$ and strong model $M_s$. Run both on
$x$ and let $a_w(x),a_s(x)\in\{0,1\}$ record whether each answer is
correct under one task evaluator: a user's judgment, a reference value,
or an executable test (Section~\ref{sec:feedback}). The names
\emph{weak} and \emph{strong} describe overall roles; neither model is
assumed correct on every query. A \emph{pre-generation} policy
$r(x)\in\{w,s\}$ selects one model before either answer is generated,
from the query and features computed from it alone; for a new query
only the selected model runs. We call a system that implements such a
policy a \emph{pre-generation router}.

\paragraph{Accuracy.}
Upgrading to strong changes accuracy only in two of the four paired
outcomes. Define the realized gain $g(x)=a_s(x)-a_w(x)$: it is $+1$ when only
strong is correct ($a_w=0,a_s=1$), the \emph{only-strong-correct} (OSC) outcome; $-1$
when strong replaces a correct answer ($a_w=1,a_s=0$); and $0$ when both
models agree in correctness. In the review filter of
Section~\ref{sec:intro}, an explicit ``the screen arrived cracked'' is
classified correctly by both models ($g=0$). A review that describes
the defect only indirectly is the OSC case ($g=+1$). A review
that the weak model classifies correctly but the strong model
over-interprets gives $g=-1$. For a workload $T$, policy accuracy
satisfies
\begin{equation}
 A_T(r)=\frac{1}{|T|}\sum_{x\in T}a_{r(x)}(x)
 =A_T(M_w)+\frac{1}{|T|}\sum_{x\in T:r(x)=s}g(x).
 \label{eq:paired-accuracy}
\end{equation}
At routing time $g(x)$ is unknown. We model a task as a distribution
over tuples $(X,a_w,a_s,c_w,c_s)$, of which the workload and the
records of Section~\ref{sec:feedback} are samples, and write the
expected gain of upgrading query $x$ as
\begin{equation}
 \mathbb E[g\mid x]
 =\Pr(a_w=0,a_s=1\mid x)
 -\Pr(a_w=1,a_s=0\mid x).
 \label{eq:ideal-gain}
\end{equation}

\paragraph{Cost.}
Let $c_w(x)$ and $c_s(x)$ be the realized answer-generation costs. Under
token pricing,
\begin{equation}
 c_j(x)=\pi_j^{\rm in}n_j^{\rm in}(x)
 +\pi_j^{\rm out}n_j^{\rm out}(x),\qquad j\in\{w,s\},
 \label{eq:token-cost}
\end{equation}
where $n_j^{\rm in}$ and $n_j^{\rm out}$ are input and output token
counts and $\pi_j^{\rm in}$ and $\pi_j^{\rm out}$ their per-token
prices. Input length follows from the prompt and tokenizer, but output
length is unknown before generation: one coding query elicits a short
function and another a long program; extraction queries return
different numbers of items. Costs therefore vary across queries and
must be estimated before generation. The total answer-model cost of a
pre-generation policy on $T$ is
\begin{align}
 C_T(r)&=\sum_{x\in T}c_{r(x)}(x) \notag\\
 &=\sum_{x\in T}c_w(x)
 +\sum_{x\in T:r(x)=s}[c_s(x)-c_w(x)].
 \label{eq:total-cost}
\end{align}
This is the cost of answering every query with weak plus the
incremental cost of the selected upgrades. Let
$C_s(T)=\sum_{x\in T}c_s(x)>0$ be the all-strong cost. We report cost as
a fraction of $C_s(T)$, so routing every query to strong has fraction
one. Because $c_s(x)-c_w(x)$ varies across queries, the fraction of
queries sent to strong is not a substitute for this cost fraction, and
we report dollars even where a policy is controlled by a call quota.

\paragraph{The budgeted routing problem.}
Given a fixed model pair and a budget fraction $B\in(0,1]$, choose a
pre-generation policy $r$ that solves
\begin{equation}
 \max_r\ \mathbb E[a_{r(X)}(X)]
 \quad\text{subject to}\quad
 \mathbb E[c_{r(X)}(X)]\le B\cdot\mathbb E[c_s(X)].
 \label{eq:ideal-routing-objective}
\end{equation}
Write $G(x)=\mathbb E[g\mid x]$ and $D(x)=\mathbb E[c_s-c_w\mid x]$. By
Lagrangian relaxation (the fractional-knapsack argument), the optimal
policy upgrades exactly the queries whose expected gain per incremental
cost $G(x)/D(x)$ exceeds a cutoff $\lambda\ge0$ (Section~\ref{sec:score-step} handles $D(x)\le0$ with a floor). A
router therefore needs two estimates from the query alone: where the
upgrade pays off, $G(x)$ (Section~\ref{sec:feedback} explains which
part we estimate), and what it costs, $D(x)$. How the cutoff is set depends on whether the
workload is known in advance (\emph{batch}) or arrives one query at a
time (\emph{streaming}); Section~\ref{sec:operating-point} treats both.

\subsection{Existing routing policies}
\label{sec:existing}
\begin{table}[t]
\centering
\caption{What each routing policy provides in the two-model setting.
$^{\dagger}$BARGAIN's guarantee concerns agreement with the strong model,
\system's expected normalized spending.}
\label{tab:baseline-features}
\footnotesize
\setlength{\tabcolsep}{3pt}
\renewcommand{\arraystretch}{1.08}
\newcommand{\ftyes}{\checkmark}
\newcommand{\ftno}{\textcolor{gray}{$\times$}}\begin{tabular}{@{}lccccc@{}}
\toprule
 & \rotatebox{60}{RouteLLM} & \rotatebox{60}{FrugalGPT} & \rotatebox{60}{BARGAIN} & \rotatebox{60}{OctoSelector} & \rotatebox{60}{\system}\\
\midrule
Decides before generation & \ftyes & \ftno & \ftno & \ftyes & \ftyes\\
Trained on the task's own labels & \ftno & \ftyes & \ftno & \ftyes & \ftyes\\
Per-query cost prediction & \ftno & \ftno & \ftno & \ftno & \ftyes\\
Small labeled sample suffices & \ftno & \ftno & \ftyes & \ftno & \ftyes\\
Batch and streaming & \ftyes & \ftyes & \ftno & \ftyes & \ftyes\\
Finite-sample guarantee & \ftno & \ftno & \ftyes$^{\dagger}$ & \ftno & \ftyes$^{\dagger}$\\
\bottomrule
\end{tabular}
\end{table}

Two families of policies address this problem with different evidence;
Table~\ref{tab:baseline-features} summarizes what each provides.
A \emph{cascade} runs the weak model first and inspects its answer
before escalating. FrugalGPT learns answer-quality scorers for budgeted
cascading \cite{chen2024frugalgpt}; BARGAIN ranks queries by the weak
model's confidence and sets a threshold from sampled oracle outputs,
with a finite-sample guarantee on agreement with the oracle
\cite{zeighami2025bargain}. The weak answer is informative evidence, but
it is bought with a weak call on every query. A cascade that makes the
same final choices as policy $r$ pays
$\sum_{x\in T}c_w(x)+\sum_{x\in T:r(x)=s}c_s(x)$, which exceeds
Eq.~\ref{eq:total-cost} by exactly the weak call on every escalated
query, plus any oracle samples used to set its threshold. It also cannot
escalate until the weak answer exists, which Section~\ref{sec:e2e}
measures as time. A pre-generation router decides from the input alone:
RouteLLM provides a preference score pretrained on human comparisons
between model pairs \cite{ong2025routellm}, OctoSelector profiles
every candidate model on a labeled training set, clusters queries by
difficulty features, and assigns queries to models with an integer
program over cluster-level estimates \cite{zhang2026octoselector}, and related routers learn
from thousands of labeled queries
\cite{ding2024hybridllm,chen2024routerdc}, far more than the feedback we
assume in Section~\ref{sec:feedback}. BARGAIN also needs the whole
batch in hand, since its threshold is a prefix of the batch ranked by
confidence, whereas the other three can also decide queries as they
arrive.

\subsection{Task alignment}
\label{sec:feedback}
Every task has its own notion of a correct answer. For the review
filter it is the user's predicate; for the value-filling tasks it is
the database value; for the coding task it is a set of tests. The
outcomes $a_w(x)$ and $a_s(x)$ of Section~\ref{sec:problem} are
defined against that notion, and so are the gain $G(x)$ and the
incremental cost $D(x)$ that a router has to estimate. We say a router
is \emph{aligned} with a task when its estimates are fitted to that
task: to its correctness criterion, to the two models actually in use,
and to the queries the workload contains.

None of the policies in Section~\ref{sec:existing} is aligned in this
sense, and none estimates $G(x)$ and $D(x)$ per query for the task
at hand from a small labeled sample. RouteLLM's preference score was trained on comparisons from tasks and model pairs unrelated to the task of interest, so it is not trained on the OSC outcome of this pair on this task.
A cascade trusts a signal about the weak answer, its confidence for
BARGAIN or a learned score of its correctness for FrugalGPT, which says
nothing about whether the strong model would do better; Section~\ref{sec:main-results} measures both signals. OctoSelector fits its estimates to the task, but only at cluster granularity, so two queries in the same cluster always receive the same estimate. On the cost side, none of them estimates the incremental cost $D(x)$ of a new query before generation: as run in Section~\ref{sec:setup}, RouteLLM and BARGAIN rank queries with no cost term, FrugalGPT budgets against the mean strong-call cost, and OctoSelector against per-cluster mean costs.
\system therefore fits its estimates to each task separately, from a
small sample of that task.

\paragraph{The sample.}
\system draws a small set of $n$ paired records from the task and
calls it $\mathcal D_{\rm tr}$. Each record
$(x_i,a_{w,i},a_{s,i},c_{w,i},c_{s,i})$ costs two model calls, whose
prices are simply observed, and two correctness judgments. The
judgments are what ties the sample to the task. In a deployment they
come from the user who owns the workload and can tell a right answer
from a wrong one; in our experiments a reference value or a test stands
in for that user.
Each judgment costs the user's time and holds for this task and
model pair only, so we keep $n$ small, far below the thousands of
labeled queries that the learned routers of Section~\ref{sec:existing}
assume ($n=50$ in our experiments). If a task provides no judgments at all, the strong model's own
answer can serve as the reference for the weak one;
Section~\ref{sec:strong-label} shows that this cruder signal recovers
most of the benefit.
From these records \system fits two predictors that act on the query
alone: one for the OSC probability $\Pr(g=+1\mid x)$, the positive term
of Eq.~\ref{eq:ideal-gain}, and one for each model's generation cost,
which needs no correctness label. We do not model the $g=-1$ outcome,
in which the strong model replaces a correct weak answer: it occurs on
0.1--5.4\% of queries in our tasks, too rarely for a small sample to
capture, so the score is a surrogate for $G(x)$ that leaves this term
out. The cost of the paired calls is charged to \system in every
comparison (Section~\ref{sec:setup}).

\subsection{Overview}
\label{sec:pipeline-overview}
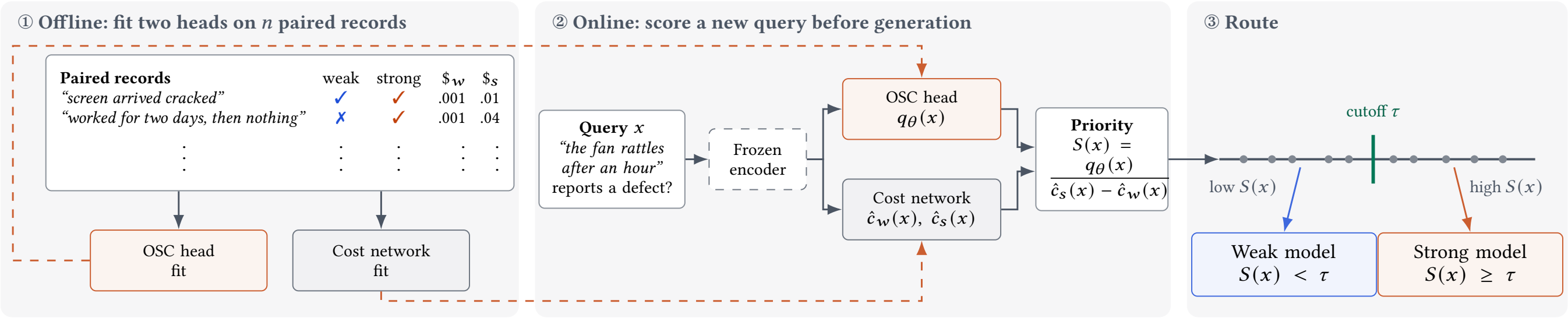
\begin{figure*}[t]
\centering
\resizebox{\textwidth}{!}{%
\begin{tikzpicture}[
 x=1cm,y=1cm,
 >={Latex[length=1.8mm]},
 flow/.style={->,semithick,draw=RRGray},
 fitarrow/.style={->,semithick,draw=RROrange!85,dashed},
 box/.style={draw=RRGray!65,rounded corners=2pt,fill=white,
  minimum height=0.75cm,align=center,inner sep=4pt,font=\footnotesize},
 frozen/.style={box,draw=RRGray!80,dashed},
 strongbox/.style={box,draw=RROrange!85,fill=RROrange!6},
 weakbox/.style={box,draw=RRBlue!85,fill=RRBlue!6},
 costbox/.style={box,draw=RRGray!80,fill=RRGray!8},
 lane/.style={draw=none,fill=RRGray!5,rounded corners=4pt},
 hdr/.style={anchor=north west,font=\footnotesize\bfseries,text=RRGray},
 lbl/.style={font=\scriptsize,text=RRGray,align=center}
]
 \fill[lane] (-0.85,-1.95) rectangle (5.55,1.95);
 \fill[lane] (5.75,-1.95) rectangle (13.6,1.95);
 \fill[lane] (13.8,-1.95) rectangle (18.5,1.95);
 \node[hdr] at (-0.75,1.9) {\ding{172} Offline: fit two heads on $n$ paired records};
 \node[hdr] at (5.85,1.9) {\ding{173} Online: score a new query before generation};
 \node[hdr] at (13.9,1.9) {\ding{174} Route};

 \node[box,inner sep=5pt,font=\scriptsize] (recs) at (2.6,0.45) {%
  \begin{tabular}{@{}l@{\hspace{6pt}}c@{\hspace{6pt}}c@{\hspace{6pt}}r@{\hspace{5pt}}r@{}}
   \multicolumn{1}{@{}l}{\textbf{Paired records}} & weak & strong & \$$_w$ & \$$_s$\\
   \textit{``screen arrived cracked''} & \textcolor{RRBlue}{\ding{51}} & \textcolor{RROrange}{\ding{51}} & .001 & .01\\
   \textit{``worked for two days, then nothing''} & \textcolor{RRBlue}{\ding{55}} & \textcolor{RROrange}{\ding{51}} & .001 & .04\\
   \multicolumn{1}{@{}c}{$\vdots$} & $\vdots$ & $\vdots$ & $\vdots$ & $\vdots$
  \end{tabular}};
 \node[strongbox,text width=1.9cm,font=\scriptsize] (fh) at (1.35,-1.25) {OSC head\\fit};
 \node[costbox,text width=1.9cm,font=\scriptsize] (fc) at (3.85,-1.25) {Cost network\\fit};
 \draw[flow] (recs.south) ++(-1.25,0) -- (fh.north);
 \draw[flow] (recs.south) ++(1.25,0) -- (fc.north);

 \node[box,text width=1.52cm,font=\scriptsize] (q) at (6.7,0.0)
  {\textbf{Query $x$}\\\textit{``the fan rattles after an hour''}\\reports a defect?};
 \node[frozen,text width=0.92cm,font=\scriptsize] (enc) at (8.5,0.0) {Frozen\\encoder};
 \node[strongbox,text width=1.67cm,font=\scriptsize] (ph) at (10.525,0.62)
  {OSC head\\$q_\theta(x)$};
 \node[costbox,text width=1.67cm,font=\scriptsize] (pc) at (10.525,-0.62)
  {Cost network\\$\hat c_w(x),\ \hat c_s(x)$};
 \node[box,text width=1.35cm,font=\scriptsize] (score) at (12.75,0.0)
  {\textbf{Priority}\\$S(x)=\dfrac{q_\theta(x)}{\hat c_s(x)-\hat c_w(x)}$};
 \draw[flow] (q) -- (enc);
 \draw[flow] (enc.east) -- ++(0.2,0) |- (ph.west);
 \draw[flow] (enc.east) -- ++(0.2,0) |- (pc.west);
 \draw[flow] (ph.east) -- ++(0.2,0) |- ([yshift=0.15cm]score.west);
 \draw[flow] (pc.east) -- ++(0.2,0) |- ([yshift=-0.15cm]score.west);
 \draw[fitarrow] (fh.west) -- ++(-0.95,0) |- (3.0,1.4) -- (10.525,1.4) -- (ph.north);
 \draw[fitarrow] (fc.south) |- (10.525,-1.75) -- (pc.south);

 \draw[flow] (score.east) -- (14.2,0.0);
 \draw[thick,draw=RRGray] (14.2,0.0) -- (18.1,0.0);
 \foreach \p in {14.5,14.85,15.1,15.55,15.8,16.35,16.6,17.0,17.35,17.7} \fill[RRGray!70] (\p,0.0) circle (1.4pt);
 \draw[very thick,draw=RRGreen] (16.1,-0.3) -- (16.1,0.3);
 \node[lbl,text=RRGreen!80!black] at (16.1,0.6) {cutoff $\tau$};
 \node[lbl] at (14.5,-0.35) {low $S(x)$};
 \node[lbl] at (17.75,-0.35) {high $S(x)$};
 \node[weakbox,text width=2.0cm] (weak) at (15.0,-1.3) {Weak model\\$S(x)<\tau$};
 \node[strongbox,text width=2.0cm] (strong) at (17.3,-1.3) {Strong model\\$S(x)\ge\tau$};
 \draw[flow,draw=RRBlue!85] (15.2,-0.1) -- (weak.north);
 \draw[flow,draw=RROrange!85] (17.1,-0.1) -- (strong.north);
\end{tikzpicture}%
}
\caption{\system overview. \ding{172} A few paired records fit an OSC head and a cost network on a frozen encoder. \ding{173} A new query is scored by its OSC probability over its predicted extra cost. \ding{174} A cutoff $\tau$ sends it to the weak or the strong model; \systemB and \systemS differ only in how $\tau$ is chosen (Section~\ref{sec:operating-point}).}
\label{fig:method-flow}
\end{figure*}

 Section~\ref{sec:method} gives the predictors and both
operating points (Figure~\ref{fig:method-flow},
Algorithm~\ref{alg:aroute}); Section~\ref{sec:experiments} compares them
with FrugalGPT, BARGAIN, RouteLLM, and OctoSelector;
Section~\ref{sec:related} surveys related work; and
Section~\ref{sec:limitations} discusses scope and future work.

\section{ARoute}
\label{sec:method}
{\color{black}
\system learns a pre-generation priority from the paired records of
Section~\ref{sec:feedback}. A shared frozen representation supports
two prediction targets: the OSC outcome and each model's
generation cost. \par}

Figure~\ref{fig:method-flow} follows one query through the pipeline. Offline, the paired records are encoded once by the frozen BERT (Section~\ref{sec:records}) and the two heads are fitted on the cached vectors (Section~\ref{sec:targets}). Online, a new query (a review, in the running example) is encoded the same way; the OSC head returns the probability that only the stronger model is correct, the cost network returns both predicted costs, their ratio gives the priority (Section~\ref{sec:score-step}), and a cutoff selects the one model that answers (Section~\ref{sec:operating-point}). Algorithm~\ref{alg:aroute} states the same pipeline in full.\par

\begin{algorithm}[t]
\caption{\system: training and routing}
\label{alg:aroute}
\begin{algorithmic}[1]
\Require paired records $\{(x_i,a_{w,i},a_{s,i},c_{w,i},c_{s,i})\}_{i=1}^{n}$; frozen BERT; budget fraction $B$ (batch) or quota $\beta$, validation size $m$, and confidence $\delta$ (streaming)
\Statex \textbf{Offline (once per task)}
\State $e_i\gets e(x_i)$ for all $i$ \Comment{frozen [CLS] embedding, Eq.~\ref{eq:embedding}}
\State $y_i\gets\mathbf 1[a_{w,i}=0,\,a_{s,i}=1]$ \Comment{OSC outcome}
\State fit linear head $q_\theta$ on $\{(e_i,y_i)\}$ by weighted BCE \Comment{Eq.~\ref{eq:head-loss}}
\State fit cost MLP $f_\phi$ on $\{(e_i,\log c_{w,i},\log c_{s,i})\}$ \Comment{epochs by 3-fold CV}
\State $\bar c_s\gets\frac1n\sum_i c_{s,i}$
\Statex \textbf{Online (per query $x$, before any generation)}
\State $e\gets e(x)$;\ $q\gets q_\theta(e)$;\ $(\hat c_w,\hat c_s)\gets\exp f_\phi(e)$
\State $S(x)\gets q\,\bar c_s/\max\{\hat c_s-\hat c_w,\,0.01\bar c_s\}$ \Comment{Eq.~\ref{eq:score}}
\State \textbf{if} $S(x)\ge\tau$ \textbf{then} call strong \textbf{else} call weak
\Statex \textbf{Choosing $\tau$}
\State \textbf{batch (\systemB):} sort the batch by $S$; the strong prefix is the longest whose realized cost, plus acquisition, fits the budget \Comment{Sec.~\ref{sec:batch-case}}
\State \textbf{streaming (\systemS):} pay $m$ paired validation calls; $\tau\gets$ smallest grid cutoff passing the KL test of Eq.~\ref{eq:share-test} at allowance $b$; re-run every 50 routed queries with the remaining allowance, and guard \Comment{Sec.~\ref{sec:streaming-case}}
\end{algorithmic}
\end{algorithm}

\Needspace*{4\baselineskip}
\subsection{\textcolor{black}{Paired records and shared representation}}
\label{sec:records}
{\color{black}
The training set $\mathcal D_{\rm tr}$ of Section~\ref{sec:feedback}
supplies the supervision for both prediction targets, and all score
fitting and tuning uses it alone; the validation records of \systemS
do not train the predictors.\par}

{\color{black}
To limit the number of trainable parameters, both predictors share a
frozen \texttt{bert-base-uncased} encoder \cite{devlin2019bert} in evaluation mode.
Long queries are divided into model-length windows so that the entire
prompt is represented. Let $h_j(x)$ be the final-layer \texttt{[CLS]}
state for window $j$ and $m_j(x)$ its non-padding token count. Both
heads use the same cached, unit-normalized 768-dimensional vector
\begin{equation}
 e(x)=\operatorname{Normalize}\!\left(
 \frac{\sum_j m_j(x)h_j(x)}{\sum_j m_j(x)}\right).
 \label{eq:embedding}
\end{equation}
No BERT weights are updated, so fitting the heads reduces to small
problems over cached vectors.\par}

\subsection{\textcolor{black}{Prediction targets}}
\label{sec:targets}
{\color{black}
The two predictors separate the potential benefit of an upgrade from
its price. The OSC head targets the positive outcome in
Eq.~\ref{eq:ideal-gain}, while the cost network predicts the two costs
needed to estimate the incremental expense in
Eq.~\ref{eq:total-cost}.\par}

\paragraph{\textcolor{black}{OSC outcome.}}
The OSC head targets $y(x)=\mathbf{1}[a_w(x)=0,\,a_s(x)=1]$, the
event that only strong is correct. A single linear sigmoid head predicts
$q_\theta(x)=\Pr_\theta(y=1\mid x)$, an estimate of $\Pr(g=+1\mid x)$.
The two classes are rarely balanced in a 50-record sample, and OSC can be the majority (at least 72\% of LiveCodeBench queries, Table~\ref{tab:datasets}) as well as the minority, so
training example $i$ receives the
class weight $w_i=n/(2n_+)$ if $y_i=1$ and $w_i=n/(2n_-)$ otherwise,
where $n_+$ and $n_-$ count positive and negative examples and
$n=n_++n_-$. With $\mathcal B$ the current mini-batch, the objective is
the weighted binary cross-entropy
\begin{equation}
 \mathcal{L}_{\rm OSC}
 =\frac{1}{|\mathcal B|}\sum_{i\in\mathcal B}w_i\ell_{\rm BCE}(q_\theta(x_i),y_i).
 \label{eq:head-loss}
\end{equation}
We train the 769-parameter head for eight epochs with AdamW \cite{loshchilov2019adamw} (learning
rate 0.01, batch size eight). The class weights shift the scale of its output, which is therefore not calibrated and enters Eq.~\ref{eq:score} only as part of the surrogate for $G(x)/D(x)$, whose value we measure empirically (Section~\ref{sec:cost-ablation}).

\paragraph{\textcolor{black}{Generation costs.}}
{\color{black}
Because output length is unknown before generation, we predict each
model's cost from the query. The same $e(x)$ feeds a cost network of
three linear layers with widths
$768\!\rightarrow\!128\!\rightarrow\!32\!\rightarrow\!2$ and ReLU
between them. Its two outputs estimate log weak and strong costs and are
exponentiated to give $\hat c_w(x)$ and $\hat c_s(x)$. Features and log
targets are standardized on training rows only; training uses full-batch
smooth-$L_1$ loss with AdamW (learning rate $10^{-3}$, weight decay
0.01). The epoch count is chosen by three-fold cross-validation
\emph{within the training records} on strong log-cost mean absolute
error, after which the network is refit on all records, so no extra
paired set is needed. Neither head receives hand-engineered
text-structure features.\par}

\subsection{\textcolor{black}{Priority score}}
\label{sec:score-step}
{\color{black}
To prioritize potential improvements per unit of expense, we divide the
OSC score by predicted incremental cost. Let $\bar c_s$ be the
mean observed strong cost on training records. We use
\begin{equation}
 d(x)=\max\{\hat c_s(x)-\hat c_w(x),0.01\bar c_s\},\qquad
 S(x)=\frac{q_\theta(x)}{d(x)/\bar c_s}.
 \label{eq:score}
\end{equation}
The floor keeps the ratio stable when the estimated incremental cost is
near zero or negative. A cutoff $\tau$ sends $x$ to strong when
$S(x)\ge\tau$ and to weak otherwise; raising $\tau$ sends more queries
to weak. Thus the cutoff controls the set of selected upgrades; the
resulting accuracy and dollar cost are evaluated empirically.\par}

\subsection{\textcolor{black}{Batch and streaming policies}}
\label{sec:operating-point}

\subsubsection{\textcolor{black}{\systemB: retrospective batch evaluation}}
\label{sec:batch-case}
{\color{black}
For a known batch of queries, \systemB keeps the strong answers of its $n$ training records and sorts the other queries by $S(x)$ and
assigns a prefix to strong and the remaining queries to weak. The
ranking uses only query features and predictors fitted on training
records. To compare rankings at equal spending, our evaluation uses
cached realized costs to select the prefix boundary retrospectively,
including the acquisition calls in total cost. When adjacent operating
points bracket the target, we use fractional assignment of the boundary
query between the two models (Section~\ref{sec:setup}).\par}

{\color{black}
This protocol compares rankings at matched observed cost; running \systemB on a live batch would need a stopping rule on realized spending, which we do not evaluate.\par}

\subsubsection{\textcolor{black}{\systemS: validation-selected streaming policy}}
\label{sec:streaming-case}
 {\color{black}
For the certificate below, let $N$ be the original workload size. We
sample $n$ training queries and $m$ additional validation queries without
replacement, pay for each paired execution once, and retain their strong
answers. The policy routes only the remaining $R=N-n-m>0$ queries as
they arrive. Let $\beta\in(0,1)$ specify the normalized-cost quota defined
below, including acquisition costs.\par}

In the streaming case, the cutoff $\tau$ must be chosen before any new
request is answered. Because the training records were already used to fit the score, we choose it on a small separate validation set of
paired costs. A finite-sample test then certifies that the chosen cutoff
keeps expected spending within a quota. The quota is stated in
\emph{normalized} units: each query's cost is divided by the cost of its
more expensive call, so every value lies in $[0,1]$. We use this unit
instead of dollars because a dollar guarantee would require a known
maximum price per call, which is rarely available and, when it is, is
too loose to be useful. Note that the quota is neither a dollar budget
nor, in general, a fraction of queries sent to the strong model. When strong is the costlier call on every query, a strong call uses one unit and a weak call $c_w/c_s$, so on our price ratios the quota is close to a quota on strong calls (Section~\ref{sec:streaming-results}).

\paragraph{Assumptions.}
The model pair, prices, prompts, and decoding are fixed, so each paired
record is a draw from one distribution. The original batch is an
i.i.d.\ sample from it, the training/validation split does not depend
on record values, and the score and cutoff grid are frozen before the
validation records are inspected. In a live stream the records are
the first $n+m$ arrivals, which satisfy these assumptions when arrival
order is independent of the records.
\paragraph{Validation data and candidate cutoffs.}
{\color{black}
The validation sample is $V=\{(X_i,c_{w,i},c_{s,i})\}_{i=1}^{m}$;
cutoff selection uses its paired costs without correctness labels. Under the assumptions above, conditional on training, the validation records are i.i.d.\ and independent of the remaining records, the frozen score $S$, and the finite grid $t_1<\cdots<t_K$, built from training data or an independent reference pool. For candidate $t$,
the policy chooses weak if $S(X)<t$ and strong otherwise, so its actual
cost on a record is
\begin{equation}
 c_t=\mathbf1[S(X)<t]\cdot c_w+\mathbf1[S(X)\ge t]\cdot c_s.
 \label{eq:actual-cost}
\end{equation}
Here and below, $c_t$ and the quantities $r_w$, $r_s$, $r_t$, $\ell_t$
defined next are functions of one paired record $(X,c_w,c_s)$.\par}

\paragraph{Normalized costs and the quota.}
\label{sec:certificate}
Let $M=\max(c_w,c_s)$ be the cost of the record's more expensive call
and define
\begin{equation}
 \begin{aligned}
 r_w&=\frac{c_w}{M},\qquad r_s=\frac{c_s}{M},\\
 r_t&=\mathbf1[S(X)<t]\cdot r_w+\mathbf1[S(X)\ge t]\cdot r_s,
 \end{aligned}
 \label{eq:share}
\end{equation}
with $r_s$ and $r_w$ equal to zero if $M=0$. Thus $r_t\in[0,1]$: an answer costing
half as much as the larger call uses half a unit. Answering all $N$ queries, one answer each,
costs at most $N$ units, so the quota is $\beta\cdot N$; it equals a
fraction $\beta$ of all-strong normalized cost only when $c_s\ge c_w>0$
almost surely. Let $F$ upper-bound the normalized cost of the $n+m$
sampled pairs, declared before validation together with $N$, $\beta$,
and $m$; each pair uses at most two units, so $F=2(n+m)$ is always
a valid upper bound for the labeled pairs. The target is
\begin{equation}
 F+R\cdot\mathbb E[r_t]\le\beta\cdot N,\qquad R=N-n-m,
 \label{eq:share-target}
\end{equation}
where $\mathbb E[r_t]$ is the population mean of the normalized cost of
one routed query, covering query variation and generation randomness
with prices, models, prompts, and decoding fixed. To compare many
cutoffs without a multiple-testing penalty, we audit the conservative
statistic
\begin{equation}
 \ell_t=r_w+\mathbf1[S(X)\ge t]\cdot(r_s-r_w)_+,
 \label{eq:share-envelope}
\end{equation}
which charges the larger normalized cost whenever strong is selected.
Hence $r_t\le\ell_t\le1$ on every record, $\ell_t$ never increases as
the cutoff rises even if weak is the more expensive answer, and $\ell_t=r_t$
whenever weak is cheaper. Its population mean is $\mu_t=\mathbb E[\ell_t]$,
an unknown number in $[0,1]$ that satisfies $\mathbb E[r_t]\le\mu_t$; the
certificate below therefore tests $\mu_t\le b$, and does so using the
observable sample mean $\bar\ell_t$.
\paragraph{The certificate.}
The allowance per remaining query is $b=(\beta\cdot N-F)/R$. If
$b\le0$, return $\perp$; if $b\ge1$, every candidate meets the allowance.
For $b\in(0,1)$, let $\bar\ell_t=m^{-1}\sum_{i=1}^m\ell_{t,i}$ be the
sample mean of $\ell_t$ over the validation records, and certify $t$ if
\begin{equation}
 \bar\ell_t<b,\qquad m\cdot D_{\mathrm{KL}}(\bar\ell_t\|b)\ge\log(1/\delta),
 \label{eq:share-test}
\end{equation}
where $D_{\mathrm{KL}}(a\|b)=a\log(a/b)+(1-a)\log((1-a)/(1-b))$ is the
Kullback--Leibler divergence between $\mathrm{Bernoulli}(a)$ and
$\mathrm{Bernoulli}(b)$, with endpoint limits, and $\delta\in(0,1)$ bounds the
probability of a false certificate. The first condition puts the
sample mean below the allowance; the second requires the gap between
them to be large enough, given $m$, that a population mean above the
allowance is unlikely to have produced it. Return the smallest certified cutoff, which uses strong on
the largest set of queries, or $\perp$ if none passes; the replays use
$\delta=0.05$.

\begin{theorem}[Normalized-cost quota]
\label{thm:share-certificate}
Fix $N$, $n$, $m$, $\beta$, $F$, $\delta$, the score, and the cutoff
grid before inspecting validation records. Under the i.i.d.-batch split
above, conditional on training, the probability over the validation
sample $V$ of returning a cutoff whose expected normalized cost exceeds
the quota is at most $\delta$:
\begin{equation}
 \Pr_V\bigl\{\hat t\ne\perp,\ F+R\cdot\mathbb E[r_{\hat t}]>\beta\cdot N\bigr\}\le\delta.
 \label{eq:share-guarantee}
\end{equation}
\end{theorem}
\begin{proof}
Assume $0<b<1$; the other cases are treated at the end.

\emph{One cutoff.} For i.i.d.\ $[0,1]$-valued variables with population
mean $\mu_t$, the KL-Chernoff bound limits how far below $\mu_t$ their
sample mean can fall~\cite{foong2022bounded}:
\[
 \Pr_V(\bar\ell_t\le q)\le\exp\{-m\cdot D_{\mathrm{KL}}(q\|\mu_t)\},\qquad q<\mu_t.
\]
If $m\cdot D_{\mathrm{KL}}(0\|b)<\log(1/\delta)$, no sample mean passes. Otherwise
there is a unique $q_\delta\in[0,b)$ with
$m\cdot D_{\mathrm{KL}}(q_\delta\|b)=\log(1/\delta)$, and since $D_{\mathrm{KL}}(q\|b)$ decreases on
$[0,b)$, passing the test is equivalent to $\bar\ell_t\le q_\delta$
(the strict condition $\bar\ell_t<b$ is implied because $q_\delta<b$).
Call a cutoff \emph{unsafe} if its population mean exceeds the
allowance, $\mu_t>b$. For an unsafe cutoff, since $\partial D_{\mathrm{KL}}(q\|\mu)/\partial\mu
=(\mu-q)/(\mu(1-\mu))>0$ for $\mu>q$, we have
$D_{\mathrm{KL}}(q_\delta\|\mu_t)\ge D_{\mathrm{KL}}(q_\delta\|b)$, so the probability that its sample mean passes is
at most $\exp\{-m\cdot D_{\mathrm{KL}}(q_\delta\|b)\}=\delta$.

\emph{Selecting among cutoffs.} The test runs on all $K$ candidates,
but no union bound is needed. By Eq.~\ref{eq:share-envelope} the envelope
$\ell_t$ is nonincreasing in $t$ for every record, so the unsafe cutoffs
($\mu_t>b$) form a prefix $t_1,\dots,t_{j_*}$ of the grid whose last
index $j_*$ is fixed by the population means, not by the sample, and
every sample mean $\bar\ell_t$ is nonincreasing in $t$ as well. If any
unsafe cutoff passes, then $\bar\ell_{t_{j_*}}\le\bar\ell_{t_j}\le
q_\delta$, so the fixed cutoff $t_{j_*}$ passes too; by the one-cutoff
bound this has probability at most $\delta$. Otherwise every returned
cutoff has $\mathbb E[\ell_{\hat t}]\le b$, hence
$\mathbb E[r_{\hat t}]\le b$ since $r_{\hat t}\le\ell_{\hat t}$
pointwise, which is Eq.~\ref{eq:share-target}.

\emph{Remaining cases.} The cases $b\le0$ (always $\perp$) and
$b\ge1$ ($\mathbb E[r_t]\le1\le b$ for every $t$) are immediate.
\end{proof}
The second condition of Eq.~\ref{eq:share-test} builds in a margin
$b-q_\delta$ between the allowance and the largest sample mean that
passes; it depends on $b$, $m$, and $\delta$: at $b=0.6$, $m=40$,
and $\delta=0.05$, a validation mean of at most $q_\delta\simeq0.408$
passes, a margin of about $0.192$ units. The guarantee holds for any
$m$, but the margin can leave every candidate uncertified, in which case
the rule returns $\perp$ and routes to weak. Dropping that condition leaves a
\emph{plug-in} rule with no margin, which keeps any cutoff whose
sample mean fits the allowance; it has no guarantee, and Section~\ref{sec:share-cert} measures how often it overspends.

\paragraph{Scope.}
The theorem covers one cutoff, selected once from the validation
sample and then held fixed, and it bounds that cutoff's expected
normalized cost; it says nothing about accuracy, dollars, or the
realized spending of one batch. A fixed cutoff also tends to leave part
of the quota unused, because the margin in Eq.~\ref{eq:share-test}
is conservative by design. To spend closer to the quota, the deployed
rule re-selects the cutoff periodically as the batch is routed. This
paced rule is a heuristic outside the theorem, described next, and
Section~\ref{sec:share-cert} validates both the fixed and the paced
rule empirically: quota adherence, dollar cost, and accuracy.

\paragraph{Heuristic pacing and guard.}
In cached replays, we update the cutoff every
50 routed queries using the remaining allowance $(\beta N-U)/Q$, where
$Q$ is the remaining query count and $U$ is the normalized cost spent
so far on acquisition and routed queries. A routed query's normalized
cost needs the price of the call not made; the replay reads it from the
cache, whereas a live system does not observe it
(Section~\ref{sec:limitations}). We reuse the validation sample and the KL rule
above; an allowance of at least one admits every candidate, while a
nonpositive allowance or no passing candidate selects weak. A guard
reserves $Q\hat r_w$ units for the remaining weak calls, using their
validation mean $\hat r_w$, and sends the tail to weak if spending plus
this reserve reaches the quota. These are heuristics, not repeated applications of
Theorem~\ref{thm:share-certificate}: the allowance is adaptive,
validation data are reused, acquisition is charged at its observed
value, and the guard's mean-cost reserve does not ensure zero
overshoot. We evaluate pacing empirically in Section~\ref{sec:share-cert}.
\section{Experiments}
\label{sec:experiments}

This section answers four questions. (Q1) Does \system improve accuracy
at matched total cost, with its labels charged
(Section~\ref{sec:main-results})? (Q2) Does the streaming rule stay
within its quota, does the certificate hold, and what does it cost in
accuracy (Section~\ref{sec:streaming-results})? (Q3) Where does the gain
come from: the cost term, the OSC head as a quality signal, the
reference answers behind its labels, or the number of labels
(Section~\ref{sec:ablations})? (Q4) How do end-to-end batch times
compare (Section~\ref{sec:e2e})?\par

\subsection{Setup}
\label{sec:setup}
\paragraph{Tasks.}
We evaluate Llama-3.1-8B-Instruct \cite{grattafiori2024llama3} (weak) and GPT-5 \cite{openai2025gpt5} (strong) on five
tasks (Table~\ref{tab:datasets}). \emph{MMLU-Pro}~\cite{wang2024mmlupro} and \emph{LiveCodeBench}~\cite{jain2025livecodebench}
are the LLMRouterBench releases~\cite{li2026llmrouterbench}. \emph{SuperGPQA}~\cite{du2025supergpqa} is 1,000 questions sampled proportionally across its 72 fields, with the official zero-shot prompt and answer extraction. \emph{Football score} and \emph{Formula~1 position} are two value-filling tasks built from SWAN databases \cite{zhao2025swan}: given a European match's date, league, and teams, or a season, race, and driver, the model recovers the final score or the classified finishing position from world knowledge, scored by exact agreement with the database value. We draw 1,000 matches and 2,000 results uniformly at random from 25,979 matches and 13,109 classified finishes.

\begin{table}[t]
\centering
\footnotesize
\setlength{\tabcolsep}{3pt}
\caption{Tasks. Records: batch size; labeled: paired training records per draw; weak/strong: all-Llama and all-GPT-5 batch accuracy (\%); all-strong: GPT-5 cost of the batch; CV: coefficient of variation of GPT-5 cost per query.}
\label{tab:datasets}

\begin{tabular}{@{}lrlrrrr@{}}
\toprule
Task & Records & Labeled & Weak & Strong & All-strong & CV\\
\midrule
MMLU-Pro & 1,001 & 50 & 46.65 & 85.71 & \$13.26 & 1.62\\
LiveCodeBench & 1,055 & 50 & 16.30 & 88.82 & \$54.92 & 1.46\\
SuperGPQA & 1,000 & 50 & 22.00 & 67.30 & \$21.33 & 1.39\\
Football score & 1,000 & 50 & 13.50 & 59.10 & \$12.63 & 0.48\\
Formula 1 position & 2,000 & 50 & 13.90 & 55.10 & \$41.35 & 0.55\\
\bottomrule
\end{tabular}
\end{table}

\paragraph{Models, prices, and records.}
Llama-3.1-8B-Instruct is served in BF16 with token log probabilities recorded; GPT-5 runs at medium reasoning effort. GPT-5 is priced at \$1.25/M input and \$10/M output tokens, Llama at a hosted-price proxy of \$0.015/M and \$0.020/M. All methods are evaluated on the same cached records: the Llama answers are our own generations on every task, and the GPT-5 answers are reused from the LLMRouterBench release on MMLU-Pro and LiveCodeBench (with its per-record dollar amounts) and are our own calls on the other three tasks (about \$75 in total). Because every method sees the same answer for a query routed to a given model, differences between methods reflect their routing decisions alone.\par

\paragraph{Baselines.}
We compare with four published systems. \emph{RouteLLM} is the official pretrained pre-generation score without task-specific training \cite{ong2025routellm}. \emph{FrugalGPT} is the official scorer and cascade \cite{chen2024frugalgpt}, which calls the weak model first; it uses the same 50 training records as \system, 20 held out for cascade validation and 30 for its two DistilBERT scorers \cite{sanh2019distilbert}. \emph{BARGAIN} is the released native-confidence selector \cite{zeighami2025bargain}, a budgeted cascade: Llama answers every query, queries are ranked by the sequence confidence of the Llama answer, a sample of $M=20$ GPT-5 answers selects the longest prefix whose agreement with GPT-5 meets the target at failure probability $\delta=0.05$, and the rest are escalated; it uses no training labels. \emph{OctoSelector} is the cluster-and-assign selector of Zhang et al.~\cite{zhang2026octoselector}: queries are described by its nine linguistic difficulty features, the 50 training records serve as its profiling set for the feature scaler, the $K$-means clustering (with $K$ by its elbow rule), and each cluster's mean accuracy and cost per model, and an integer program decides per cluster how many queries go to GPT-5 at a required expected accuracy.

\paragraph{Batch, labels, and answer reuse.}
A batch is every record of a task (Table~\ref{tab:datasets}). Each
experiment is repeated over five \emph{training draws}, a draw being a random 50-record training set from a 100-record pool; training and non-training queries never share an identical normalized text, and the cost network's epoch count is cross-validated inside the training records alone. The labeled records are not free: their paired weak and strong calls count toward the batch's cost. In return their strong answers are kept, so \system, FrugalGPT, and OctoSelector route only the remaining queries, whereas RouteLLM and BARGAIN use no labels and route the whole batch. A budget is a fraction of the batch's all-strong cost; total cost is the paired calls plus every routed call, including the cascades' weak calls and escalations; accuracy is counted over the whole batch, kept answers included. Encoder computation, router training, and human labeling are not charged.

\paragraph{Operating points and repetition.}
A \emph{matched point} is the accuracy at 30\%, 50\%, or 70\% of the batch's all-strong cost. No method is handed this target: each is run across the range of its own knob (the number of top-ranked queries sent to GPT-5 for \system and RouteLLM, the budget parameter for FrugalGPT, the agreement target for BARGAIN, the required expected accuracy for OctoSelector), each setting is placed at the total cost it actually incurs, and we interpolate between the two policies whose costs bracket the target. No test-set correctness label is used to pick a policy. \system reports means and standard deviations over the five training draws; on the benchmarks each draw also averages five \emph{optimizer seeds}, so the SD is across draw means. Baselines use one seed per draw.

\paragraph{Streaming replays and quotas.}
Figure~\ref{fig:deploy} adds streaming replays, which select the cutoff before routing (Section~\ref{sec:streaming-case}; protocol in Section~\ref{sec:streaming-results}). A \emph{validation draw} is one sample of $m$ validation queries from the held-out part of the batch. Quotas are in units of normalized cost (Eq.~\ref{eq:share}): one unit is a query answered at its costlier call, so a 70\% quota is $0.7N$ units for a batch of $N$ queries.

\paragraph{Reproducibility.} Code, cached model outputs, per-draw results, and the replay scripts will be released.\par

\subsection{Main comparison}
\label{sec:main-results}

\textbf{\systemB leads at every matched cost.} Table~\ref{tab:frontier}
reports batch accuracy at matched total cost under the protocol of
Section~\ref{sec:setup}, with the 50 training records charged
(2.4--6.8\% of the all-strong batch cost across tasks and draws). \systemB has the highest mean
accuracy in all 15 task--cost settings (the \systemS column is discussed in Section~\ref{sec:streaming-results}). At 70\%, its margin over the
strongest baseline is 3.57 (MMLU-Pro), 6.85 (LiveCodeBench), 0.25
(SuperGPQA), 0.58 (football score), and 4.16 (Formula~1 position)
points. Per draw, \systemB beats the strongest baseline in
59 of 75 task--cost--draw comparisons.
\begin{table*}[t]
\centering
\caption{Batch accuracy (\%) at 30\%, 50\%, and 70\% of the all-strong batch cost (rows), with dollars spent in parentheses; mean $\pm$ SD over five training draws where it varies. Best per row in bold. \systemS sets its cutoff before routing from 40 validation records at the same quotas and is shown at the dollars it actually spends.}
\label{tab:frontier}
\footnotesize
\setlength{\tabcolsep}{3pt}
\begin{tabular}{lrrrrrr}
\toprule
Task & RouteLLM & FrugalGPT & BARGAIN & OctoSelector & \systemB & \systemS\\
\midrule
MMLU-Pro & 56.14 (\$3.98) & \(58.95\pm1.99\) (\$3.98) & 55.23 (\$3.98) & \(58.99\pm1.22\) (\$3.98) & \(\mathbf{61.40\pm0.90}\) (\$3.98) & \(57.36\pm0.43\) (\$3.10)\\
 & 62.26 (\$6.63) & \(66.68\pm3.03\) (\$6.63) & 61.87 (\$6.63) & \(66.41\pm1.26\) (\$6.63) & \(\mathbf{70.18\pm1.09}\) (\$6.63) & \(64.46\pm0.73\) (\$5.12)\\
 & 69.94 (\$9.28) & \(73.36\pm2.61\) (\$9.28) & 71.29 (\$9.28) & \(73.84\pm1.39\) (\$9.28) & \(\mathbf{77.42\pm0.54}\) (\$9.28) & \(71.93\pm0.75\) (\$7.48)\\
LiveCodeBench & 30.05 (\$16.5) & \(35.77\pm4.49\) (\$16.5) & 29.81 (\$16.5) & \(41.41\pm3.73\) (\$16.5) & \(\mathbf{46.96\pm2.48}\) (\$16.5) & \(37.00\pm1.18\) (\$12.1)\\
 & 38.58 (\$27.5) & \(49.68\pm5.93\) (\$27.5) & 40.24 (\$27.5) & \(54.75\pm2.32\) (\$27.5) & \(\mathbf{62.34\pm3.09}\) (\$27.5) & \(51.58\pm1.13\) (\$21.1)\\
 & 54.60 (\$38.4) & \(64.39\pm4.94\) (\$38.4) & 54.50 (\$38.4) & \(68.06\pm2.69\) (\$38.4) & \(\mathbf{74.91\pm2.44}\) (\$38.4) & \(66.41\pm0.68\) (\$31.9)\\
SuperGPQA & 31.75 (\$6.40) & \(35.23\pm2.38\) (\$6.40) & 32.88 (\$6.40) & \(37.17\pm1.28\) (\$6.40) & \(\mathbf{38.01\pm1.19}\) (\$6.40) & \(34.46\pm1.53\) (\$5.25)\\
 & 39.50 (\$10.7) & \(43.26\pm1.62\) (\$10.7) & 40.19 (\$10.7) & \(45.97\pm1.46\) (\$10.7) & \(\mathbf{46.99\pm1.90}\) (\$10.7) & \(42.93\pm2.11\) (\$9.12)\\
 & 48.18 (\$14.9) & \(52.01\pm0.95\) (\$14.9) & 48.48 (\$14.9) & \(54.47\pm1.45\) (\$14.9) & \(\mathbf{54.72\pm1.52}\) (\$14.9) & \(52.00\pm1.83\) (\$13.5)\\
\midrule
Football score & 24.60 (\$3.79) & \(27.88\pm4.53\) (\$3.79) & 22.21 (\$3.79) & \(32.71\pm1.71\) (\$3.79) & \(\mathbf{35.76\pm2.50}\) (\$3.79) & \(30.54\pm1.03\) (\$3.04)\\
 & 33.20 (\$6.32) & \(37.85\pm5.97\) (\$6.32) & 28.18 (\$6.32) & \(42.24\pm2.96\) (\$6.32) & \(\mathbf{44.42\pm1.93}\) (\$6.32) & \(40.89\pm1.71\) (\$5.55)\\
 & 41.50 (\$8.84) & \(46.30\pm5.93\) (\$8.84) & 35.39 (\$8.84) & \(49.88\pm1.51\) (\$8.84) & \(\mathbf{50.46\pm1.60}\) (\$8.84) & \(48.99\pm1.52\) (\$8.25)\\
Formula 1 position & 23.84 (\$12.4) & \(23.19\pm1.77\) (\$12.4) & 22.51 (\$12.4) & \(26.28\pm1.48\) (\$12.4) & \(\mathbf{38.18\pm1.62}\) (\$12.4) & \(31.87\pm0.62\) (\$9.43)\\
 & 32.10 (\$20.7) & \(29.76\pm3.05\) (\$20.7) & 27.36 (\$20.7) & \(35.56\pm1.02\) (\$20.7) & \(\mathbf{43.50\pm0.67}\) (\$20.7) & \(40.83\pm0.80\) (\$17.4)\\
 & 40.55 (\$28.9) & \(37.59\pm4.18\) (\$28.9) & 35.83 (\$28.9) & \(44.00\pm0.68\) (\$28.9) & \(\mathbf{48.16\pm0.53}\) (\$28.9) & \(46.77\pm0.39\) (\$26.5)\\
\bottomrule
\end{tabular}
\end{table*}

\textbf{Why the baselines lose.} Three of the four baselines lack at least one of the two
signals motivated in Section~\ref{sec:intro}, where only GPT-5 is
correct on this task and what the upgrade costs for this query; the fourth has both, but only at cluster resolution.
RouteLLM's released preference score carries little information about
the OSC outcome (AUROC 0.45--0.61 across the five tasks, against
0.58--0.69 for the \system head fitted on 50 records) and ignores cost.
BARGAIN ranks by Llama's sequence confidence, which marks where Llama
may fail rather than where GPT-5 would succeed, so it escalates queries
both models miss; these are common on the SWAN tasks
(Table~\ref{tab:datasets}), where both label-free baselines trail
\system by 8--16 points. FrugalGPT fits two DistilBERT scorers on 30
records, which is unstable across draws, and its cascade decides without
a cost estimate. OctoSelector estimates both signals per cluster from
the same 50 records, but four to seven clusters give every query in a
cluster the same estimate, so the queries sent to GPT-5 inside a cluster
are chosen at random; fixing its $K$ at 3, 10, or 15 instead moves its
accuracy by at most 2 points and leaves \system ahead in every row. The
gap to \system is widest on LiveCodeBench, where per-query cost
prediction matters most, and on Formula~1 position, where the OSC head
alone leads OctoSelector by 4.3 points at 70\%; Section~\ref{sec:ablations}
separates the two signals.\par

\subsection{Streaming: quota adherence and guarantee}
\label{sec:streaming-results}
\begin{figure*}[t]
\centering
\includegraphics[width=0.92\textwidth]{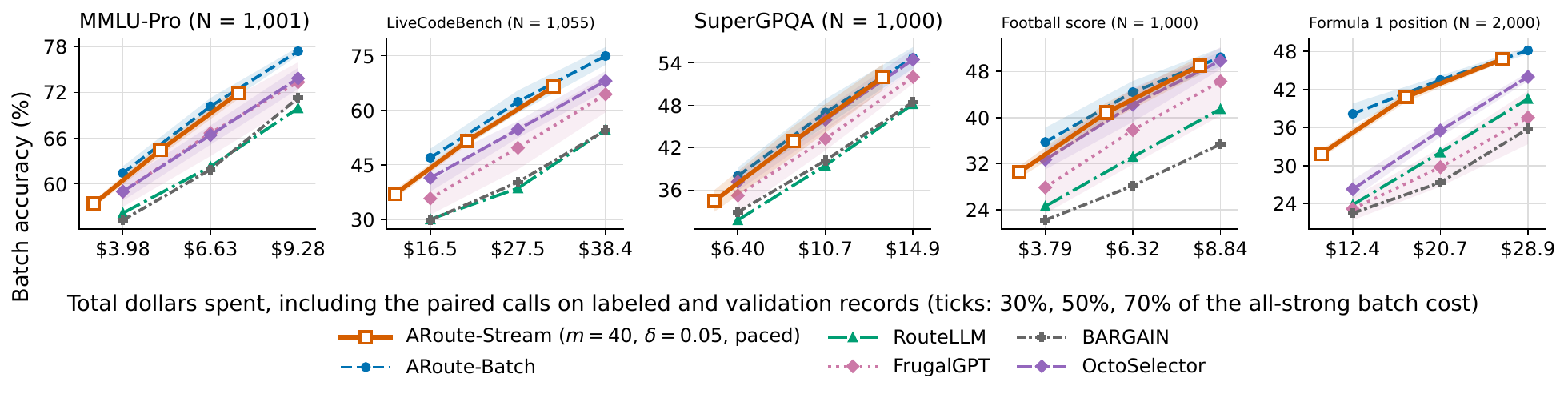}
\caption{Accuracy versus total dollars spent. Lines with bands: matched points of Table~\ref{tab:frontier} ($\pm$1 SD). Thick line: \systemS (KL test on $m=40$ validation queries, $\delta=0.05$, paced) at 30\%/50\%/70\% quotas, plotted at its realized dollars.}
\label{fig:deploy}
\end{figure*}

\paragraph{Protocol.}
The matched points of Table~\ref{tab:frontier} select each cutoff
retrospectively from the realized costs of the batch. Here we replay
\systemS on the same batches. Each replay draws $m=40$ validation
queries from the held-out part of the batch, pays their paired calls,
keeps their strong answers, certifies a cutoff with the KL test at
$\delta=0.05$ over the score quantiles of the unlabeled pool, and routes
the remaining queries in a random order, re-selecting the cutoff every
50 routed queries with the guard of Section~\ref{sec:streaming-case}.
Quotas are 30\%, 50\%, or 70\% of $N$ normalized-cost units
(Section~\ref{sec:setup}), with the labeled and validation records
charged; on our price ratios a Llama answer uses only a small fraction
of a unit, so the quota behaves nearly like a cap on strong calls. We
run 200 validation draws per trained router and report means.

\paragraph{Quota adherence.}
The paced rule hits its quota: the mean quota used is within 0.6 points
of target at all three quotas, the quota is exceeded in 2--7\% of
validation draws and never by more than 0.2 points, and the all-weak
fallback occurs in at most 1\% of draws. Dollar spending is below the
quota used, 56--65\% of the all-strong cost at the 70\% quota,
because the cost-aware score sends the
costlier queries to Llama. The \systemS column of
Table~\ref{tab:frontier} lists these accuracies and dollars next to the
matched points.

\paragraph{Accuracy at equal dollars.}
Figure~\ref{fig:deploy} plots accuracy against realized dollars. At the
50\% and 70\% quotas, \systemS exceeds every baseline's interpolated
line on all five tasks, by 0.2--8.9 points, and leads the same draw's
strongest baseline in 36 of the 50 draw--quota settings. At the 30\% quota
its spending falls below the lowest matched baseline point, so no lead
is claimed there. Interpolating the concave baselines favors \systemS;
direct matching at the replay's cost could narrow the lead.

\label{sec:share-cert}
\paragraph{Certificate validity.}
We separate the parts of the rule at the 70\% quota ($m=40$, 200
validation draws per router). Only the fixed rule tests Theorem~\ref{thm:share-certificate}
as stated, with the KL test applied once before routing: it exceeds the
quota in at most 1\% of the draws, by at most 0.2 points, and leaves the
test's margin unspent, using 51--53\% of $N$ units against the 70\%
quota. The paced rule re-applies the test every 50 queries with an
adaptive allowance and reused validation records, so their over-quota
rates are realized overspend frequencies, not the false-certificate
probability the theorem bounds; pacing recovers the unspent quota and
the accuracy of Figure~\ref{fig:deploy} at those rates. The plug-in
rule drops the KL condition: it reaches nearly the same accuracy but
exceeds the quota in 36--51\% of the draws, as an unbiased estimate
that targets the quota exactly must about half the time, whereas the KL
margin brings that rate down to 2--4\%.\par

\subsubsection{Validation-set size}
\label{sec:val-size}

\begin{figure}[t]
\centering
\includegraphics[width=0.9\columnwidth]{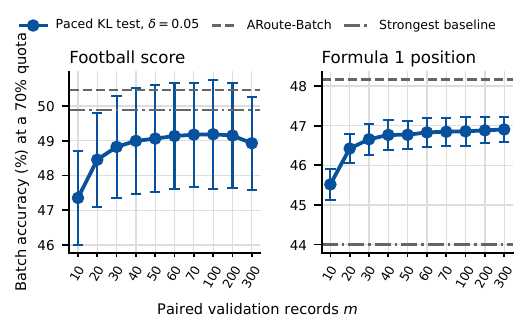}
\caption{Accuracy at a 70\% quota versus validation records $m$ for the paced KL test at $\delta=0.05$. Horizontal lines: \systemB and the best baseline matched at 70\% cost. Bars: $\pm$1 SD.}
\label{fig:val-size}
\end{figure}

Validation calls count toward the quota and their strong answers are
reused, so increasing $m$ trades routed queries for paid ones. We vary
$m$ from 10 to 300 at the 70\% quota under the protocol above.
Figure~\ref{fig:val-size} shows the two SWAN tasks; the benchmarks
behave alike, and removing the guard changes accuracy by at most 0.03
points. Accuracy rises with $m$, but slowly, and part of the rise is
mechanical, since the validation queries are answered by GPT-5 and
counted in batch accuracy: from $m=40$ to the best $m$ the gain is at
most 0.5 points, within one replay SD. Larger validation sets raise the
over-quota rate of the paced heuristic from 1--3\% at $m=10$ to 15--23\%
at $m=300$, because the KL margin shrinks as $m$ grows and the paced
rule moves toward the plug-in rule; since accuracy gains little beyond
$m=40$, a small validation set suffices. At its best $m$ the paced rule
spends 58--66\% of the all-strong dollars, less than the baselines
matched at 70\%, and with that smaller spend exceeds the strongest of
them by 2.9 points on Formula~1 position and trails it by 0.7--2.2
points on the other four tasks.

\subsection{Where does the gain come from?}
\label{sec:ablations}

\paragraph{Protocol.}
The cost-awareness ablation (Table~\ref{tab:cost-awareness-main}) uses the main protocol of Section~\ref{sec:setup}: the 50 BF16 paired records, the full-record batch, and five training draws on all five tasks, with five optimizer seeds per draw on the benchmarks. Budgets are fractions of the batch's all-strong cost, the paired calls on the training records are charged, their strong answers are reused, and accuracy is over the whole batch.

\subsubsection{Controlled cost-awareness ablation}
\label{sec:cost-ablation}
The score of Eq.~\ref{eq:score} divides the OSC probability by the predicted incremental cost, so its value over the probability alone depends on how much GPT-5's cost varies across queries. We hold each trained linear OSC head fixed and compare three scores: its probability alone, the probability divided by predicted incremental cost (the \system score of Eq.~\ref{eq:score}), and the probability divided by the observed increment. The last arm is a cost-information oracle, not an accuracy upper bound: its numerator still makes errors. Both cost arms keep the denominator floor of Eq.~\ref{eq:score}. All three arms are matched at 30\%/50\%/70\% of the batch's all-strong cost (Table~\ref{tab:cost-awareness-main}).

\begin{table}[t]
\centering
\footnotesize
\caption{Cost-awareness ablation: batch accuracy (\%) with the OSC head fixed and the cost denominator removed, predicted, or replaced by the observed increment.}
\label{tab:cost-awareness-main}
\begin{tabular}{llrrr}
\toprule
Task & Score & 30\% & 50\% & 70\%\\
\midrule
MMLU-Pro & No cost predictor & 56.71 & 64.70 & 71.62\\
 & Predicted cost & 61.40 & 70.18 & 77.42\\
 & Oracle cost & 70.42 & 79.51 & 83.56\\
LiveCodeBench & No cost predictor & 32.93 & 43.86 & 56.34\\
 & Predicted cost & 46.96 & 62.34 & 74.91\\
 & Oracle cost & 71.34 & 81.96 & 86.96\\
SuperGPQA & No cost predictor & 35.11 & 43.35 & 51.38\\
 & Predicted cost & 38.01 & 46.99 & 54.72\\
 & Oracle cost & 51.70 & 60.62 & 64.61\\
Football score & No cost predictor & 33.66 & 45.19 & 53.07\\
 & Predicted cost & 35.76 & 44.42 & 50.46\\
 & Oracle cost & 43.86 & 49.56 & 54.26\\
Formula 1 position & No cost predictor & 39.59 & 44.30 & 48.29\\
 & Predicted cost & 38.18 & 43.50 & 48.16\\
 & Oracle cost & 47.62 & 51.18 & 53.08\\
\bottomrule
\end{tabular}
\end{table}

In Table~\ref{tab:cost-awareness-main}, the gain from the cost denominator is large on the benchmarks and small or negative on the SWAN tasks. On the three benchmarks, where GPT-5's cost per query has a coefficient of variation of 1.4--1.6 (Table~\ref{tab:datasets}), predicted cost improves over the OSC score alone by 3.3--18.6 points at 70\% cost, and oracle cost adds a further 6.1--12.1; with the head fixed, this second gap isolates the potential benefit of more accurate cost estimates. The cost term is trained on paired costs alone, which the paired calls record without any correctness label, so this gain comes from the same 50 records at no added labeling cost.

On the two SWAN tasks GPT-5's cost varies less (coefficient of variation 0.5), and the cost network captures little of the variation that remains: the predicted increment adds noise rather than information, and the full score trails the head alone by up to 2.6 points; \system keeps the denominator on every task, and choosing per task from the training records whether to use it is left to future work (Section~\ref{sec:limitations}). Oracle cost still gains 3.8--4.9 points over predicted cost at 70\% cost (8.1--9.4 at 30\%), so better cost estimates would help even here.\par

\subsubsection{Quality signal at a matched strong-call rate}
\label{sec:quality-signal}
The cost denominator decides which upgrades are affordable; the numerator decides which are worth buying. To isolate the numerator, we fix the number of strong calls instead of the dollars: the 70\% of held-out queries ranked highest by a quality signal go to GPT-5, the rest to Llama, and we report held-out accuracy, so no cost term enters. This approximates the regime of the normalized-cost quota of Section~\ref{sec:streaming-case}, which on our price ratios is close to a quota on strong calls (Section~\ref{sec:streaming-results}). We compare a random order, RouteLLM's pretrained preference score, the OSC head of \system, and the true OSC indicator as an oracle (Table~\ref{tab:quality-signal}).

\begin{table}[t]
\centering
\footnotesize
\setlength{\tabcolsep}{4pt}
\caption{Held-out accuracy (\%) when the 70\% of held-out queries ranked highest by each quality signal go to GPT-5; no cost term.}
\label{tab:quality-signal}
\begin{tabular}{lrrrr}
\toprule
Task & \multicolumn{1}{c}{\shortstack[c]{Random\\order}} & \multicolumn{1}{c}{\shortstack[c]{RouteLLM\\score}} & \multicolumn{1}{c}{\shortstack[c]{OSC\\head}} & \multicolumn{1}{c}{\shortstack[c]{Oracle}}\\
\midrule
MMLU-Pro & 74.03 & 75.92 & \textbf{76.77} & 86.50\\
LiveCodeBench & 66.84 & 67.06 & \textbf{70.17} & 86.23\\
SuperGPQA & 53.61 & \textbf{57.73} & 56.27 & 70.96\\
Football score & 45.28 & 43.09 & \textbf{51.81} & 62.13\\
Formula 1 position & 42.84 & 41.41 & \textbf{47.09} & 57.93\\
\bottomrule
\end{tabular}
\end{table}

The OSC head exceeds a random order by 2.7--6.5 points on every task and RouteLLM's score by 0.9--8.7 on all tasks but SuperGPQA, where the pretrained score leads by 1.5 points. On the SWAN tasks the pretrained score falls below a random order, whereas the head gains 4.3--6.5 points: fifty task-specific labels tell the router where GPT-5 would do better, which the generic preference score does not on these tasks. The oracle column shows 10--16 points of headroom at this strong-call rate, so a better quality signal would pay off on every task.

\subsubsection{Strong model as the label oracle}
\label{sec:strong-label}
The OSC label needs a reference answer for each training record. When none is available, the strong model's own answer can stand in as the oracle: the pseudo label is then ``the weak answer differs from the strong answer'', which treats the strong model as correct and counts every disagreement as an OSC case. On MMLU-Pro and SuperGPQA the answers compared are option letters and on the SWAN tasks exact-match values; on LiveCodeBench we run both programs on the problem's test inputs and label a record as disagreeing when their outputs differ on any test, so no expected output is used. Everything else follows Table~\ref{tab:frontier}: the same 50 paired records, draws, seeds, and cost network, the paired calls still charged, and accuracy still measured against ground truth.

\begin{table}[t]
\centering
\footnotesize
\caption{Batch accuracy (\%) of \systemB when the OSC head is trained on ground-truth labels (GT) or on GPT-5's own answers as the reference (GPT-5). Agree.: share of records on which the two labels coincide.}
\label{tab:strong-label}
\setlength{\tabcolsep}{3.5pt}
\begin{tabular}{lrrrrrrr}
\toprule
 & \multicolumn{2}{c}{30\%} & \multicolumn{2}{c}{50\%} & \multicolumn{2}{c}{70\%} & \\
\cmidrule(lr){2-3}\cmidrule(lr){4-5}\cmidrule(lr){6-7}
Task & GT & GPT-5 & GT & GPT-5 & GT & GPT-5 & Agree.\\
\midrule
MMLU-Pro & 61.40 & 61.46 & 70.18 & 69.98 & 77.42 & 77.26 & 75\%\\
LiveCodeBench & 46.96 & 47.74 & 62.34 & 63.11 & 74.91 & 75.93 & 87\%\\
SuperGPQA & 38.01 & 38.09 & 46.99 & 47.44 & 54.72 & 55.39 & 75\%\\
Football score & 35.76 & 33.60 & 44.42 & 42.70 & 50.46 & 49.07 & 64\%\\
Formula 1 position & 38.18 & 37.66 & 43.50 & 42.99 & 48.16 & 47.92 & 59\%\\
\bottomrule
\end{tabular}
\end{table}

In Table~\ref{tab:strong-label}, training on GPT-5's answers changes accuracy by at most 1.0 point on the three benchmarks and by 0.2--2.2 points on the SWAN tasks, and \systemB trained this way still has the highest mean accuracy in 14 of the 15 task--cost settings. Fifty strong-model answers, which the paired calls produce anyway, thus recover most of what fifty reference answers provide.

\subsubsection{Training-set size}
\label{sec:scaling}

This study uses its own splits and reports test-set accuracy, so its numbers are not comparable with Table~\ref{tab:frontier}. Each of five split seeds divides a task's deduplicated records 40/60 into a training pool and a test set. Nested sets of 10 to 300 paired records from the pool train both predictors, with five optimizer seeds per split, and the router routes the test set at 50\% or 70\% of its all-strong cost. Each router is evaluated twice: with the paired calls on the labeled records paid from that budget, and with those calls not charged. We show the two SWAN tasks; the benchmarks follow the same pattern with larger label costs.

\begin{figure}[t]
\centering
\includegraphics[width=0.9\columnwidth]{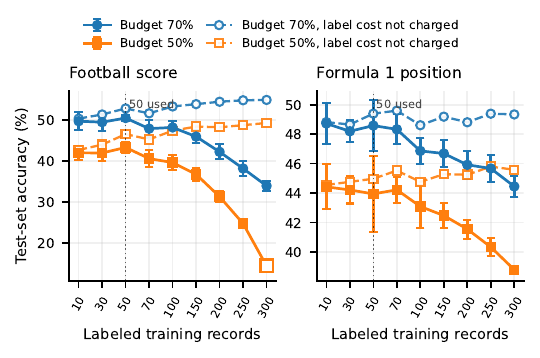}
\caption{Test-set accuracy versus labeled training records at 50\% and 70\% budgets. Solid: training-label cost charged (bars: SD over five splits); dashed: not charged.}
\label{fig:label-scaling}
\end{figure}

Figure~\ref{fig:label-scaling} shows the two curves. Without label cost, more records help modestly: from 10 to 300 records accuracy rises by 4.6 points on football score and 0.5 on Formula~1 position at the 70\% budget. With label cost charged, each labeled record spends part of the budget on a strong call chosen at random rather than by the router, and the curve turns down: 300 records cost half of the test set's all-strong cost on football score and a quarter on Formula~1 position, and accuracy falls by 16 and 4 points. Fifty records sit where the two curves separate: below 50 the predictors are still improving, above 50 they gain at most 2.1 points while the charged curve loses 4--17, and at both budgets 50 records are within 0.5 points of the best charged size. The paired calls for 50 records cost 2.6--5.9\% of the main batches (Section~\ref{sec:main-results}), which supports the setting of Section~\ref{sec:feedback}.\par

\subsection{End-to-end time}
\label{sec:e2e}

End-to-end time per batch has three parts. \textbf{Setup.} \system and FrugalGPT pay once per task for the paired calls on their 50 labeled records plus training; RouteLLM and BARGAIN use no labeled records and have no setup stage.
\textbf{Routing.} Routing is per-query CPU work, about three minutes or less per batch for every method.\footnote{\system encodes the 50 queries and fits both heads in under 10~s on an Apple M2 Pro; FrugalGPT trains two DistilBERT scorers in 25--36~s. Routing costs \system 15--147~ms per query, FrugalGPT 30~ms per Llama answer, RouteLLM one pass of its XLM-R classifier \cite{conneau2020xlmr}, BARGAIN one pass over log probabilities, and OctoSelector one feature pass and a cluster lookup.}
\textbf{Generation.} Pre-generation routers run the Llama and GPT-5 stages concurrently; cascades run them sequentially, with their GPT-5 workload scaled to the GPT-5 spending at the matched cost.

All times are measured.
Each Llama query set is submitted to vLLM 0.8.4 \cite{kwon2023vllm} (BF16) on one A40 GPU in a single call, whose wall time is the stage time; GPT-5 stages are the wall time of the selected calls issued with 16 concurrent requests. Times are taken at the 70\% point of Table~\ref{tab:frontier} and averaged over five training draws.

\begin{figure}[t]
\centering
\includegraphics[width=0.9\columnwidth]{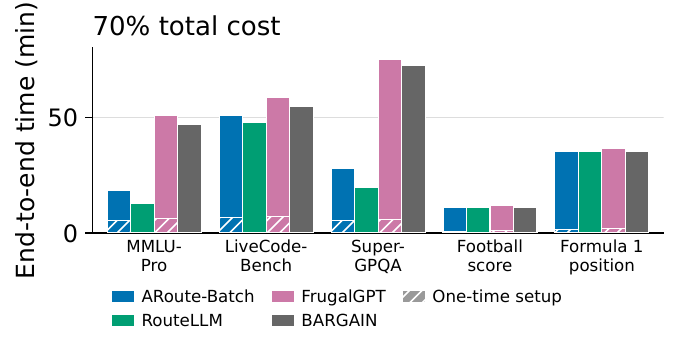}
\caption{End-to-end time per batch at 70\% total cost. Hatched: one-time setup.}
\label{fig:e2e}
\end{figure}

Figure~\ref{fig:e2e} compares end-to-end times at 70\% cost; OctoSelector, whose routing step is one feature pass and a cluster lookup, is not timed.
Where the GPT-5 stage dominates, on LiveCodeBench and the two SWAN tasks, all methods buy similar GPT-5 tokens at matched cost and are close, and \system is the fastest once its one-time setup is excluded.
Where Llama's long answers dominate, the cascades lose: on MMLU-Pro and SuperGPQA their end-to-end times are 2.5--4.0 times those of the pre-generation routers (2.5--2.8 times \system's), because BF16 Llama keeps generating until its length limit on 12\% and 24\% of the batch, respectively, and a cascade waits for every such answer.
The same answers also slow \system: a long Llama answer adds almost nothing to the dollar cost, so cost-aware routing sends many such queries to Llama, and \system is slower than RouteLLM on these two tasks and, with its setup counted, than answering every query with GPT-5.

\section{Related Work}
\label{sec:related}

\paragraph{Language models in data systems.}
Semantic operators place model calls inside filtering, joining, mapping, and aggregation pipelines, and their engines reduce the cost of those calls with proxy cascades that carry accuracy guarantees \cite{patel2025lotus}, per-operator model and prompt search on a few validation examples \cite{liu2025palimpzest,russo2025abacus}, query-engine techniques and synthesized extraction code \cite{dai2024uqe,arora2023evaporate}, accuracy-oriented plan rewrites \cite{shankar2025docetl}, and budgeted model selection \cite{huang2025thriftllm,jo2025sparellm}; SemBench and SWAN provide workloads \cite{lao2026sembench,zhao2025swan}. A longer line treats expensive models as query operators to be optimized, from specialized video cascades \cite{kang2017noscope} and prompted data wrangling \cite{narayan2022wrangle} to model-driven planners and executors over text and tables \cite{urban2024caesura,urban2024eleet,saeed2024galois,biswal2025tag,anderson2025aryn,lin2025zendb,sun2025quest,fan2024zeronl2sql}. These systems fix a model or plan per operator, document, or sub-task, or cascade per record after a proxy has run; \system supplies the per-query choice between two models that any of them can call before either model generates.

\paragraph{Pre-generation routing.}
A pre-generation router selects the answer model from the input alone: RouteLLM learns a preference score from human comparisons \cite{ong2025routellm}, Hybrid LLM predicts the quality gap between a small and a large model from about 10,000 paired comparisons \cite{ding2024hybridllm}, a conformal gate predicts from the input whether the cheap model fails where the expensive one succeeds, the OSC outcome, on thousands of labeled queries \cite{uddin2026conformalrouting}, Zooter distills reward-model scores \cite{lu2024zooter}, RouterDC and UniRoute learn query and model embeddings \cite{chen2024routerdc,jitkrittum2026uniroute}, GraphRouter predicts quality and cost on a task--query--model graph \cite{feng2025graphrouter}, FORC assigns inputs to the cheapest adequate model \cite{sakota2024forc}, and OctoSelector profiles every candidate on 7,000--9,400 training queries and assigns batches of queries to models by an integer program \cite{zhang2026octoselector}. Other routers predict each model's correctness from benchmark results \cite{shnitzer2023routing}, item response theory \cite{song2025irtrouter}, or learned model embeddings \cite{zhuang2025embedllm}, match prompts to models in a contrastively trained embedding space \cite{shirkavand2025cscr}, retrieve similar labeled examples to choose between a small and a large model \cite{lee2024orchestrallm}, or also choose how many responses the selected model samples \cite{ding2025bestroute}. Smoothie needs no labels but scores the outputs of every candidate, so it routes after generation \cite{guha2024smoothie}; LLM-Blender fuses the answers of several models \cite{jiang2023llmblender}, and Router-R1 trains an LLM to call models over several rounds \cite{zhang2025routerr1}. An empirical study finds routing promising but not feasible in every setting \cite{srivatsa2024harnessing}. CARROT proves that a router predicting per-query accuracy and cost is minimax rate-optimal \cite{somerstep2025carrot}; \system follows this plug-in design with an OSC head in place of per-model accuracies. RouterBench defines the cost--quality protocol \cite{hu2024routerbench}, and LLMRouterBench extends it and supplies the MMLU-Pro and LiveCodeBench records we use \cite{li2026llmrouterbench}; a survey organizes the literature by pre- versus post-generation timing \cite{varangot2025survey}. These routers treat cost as a per-model constant, learn from thousands of labels, or run several models per query; \system predicts each model's cost per query and fits both heads from a small set of paired records.

\paragraph{Cascades and online routing.}
A cascade runs the weak model first and escalates on a signal from its output: learned quality scorers \cite{chen2024frugalgpt}, token-level uncertainty \cite{gupta2024cascades}, self-verification \cite{aggarwal2024automix}, answer consistency \cite{yue2024mot}, or online no-regret rules \cite{nie2024onlinecascade}; theory characterizes when confidence-based deferral suffices \cite{jitkrittum2023confidence} and unifies routing with cascading \cite{dekoninck2025unified}. In data processing, BARGAIN selects cascade thresholds from sampled oracle outputs with guarantees \cite{zeighami2025bargain}, Task Cascades vary model and operation per stage \cite{shankar2026taskcascades}, and streaming cascades execute per partition \cite{liskowski2026streamingcascades}. Every cascade pays for and waits on the weak call on every input; \system decides before generation, so the two stages run concurrently (Section~\ref{sec:e2e}). Bandit routers such as MixLLM, PILOT, MetaLLM, and LLM Bandit learn from online feedback under a budget \cite{wang2025mixllm,panda2025pilot,nguyen2024metallm,li2025llmbandit} but need a stream of outcomes and give no finite-sample statement about spending.

\paragraph{Statistical guarantees and cost prediction.}
Database systems certify approximate results from proxies and samples, selecting proxy thresholds or learned filters for precision or recall targets \cite{kang2020supg,lu2018probabilisticpredicates}, bounding aggregates with expensive predicates \cite{kang2021abae}, and giving a-priori error guarantees from pilot samples \cite{zhu2025pilotdb,jo2024thalamusdb}. Selective classification, learning to defer, conformal risk control, and Learn-then-Test control a risk from a calibration set \cite{geifman2017selective,madras2018defer,angelopoulos2024crc,angelopoulos2025ltt}, recently for the error rate of routed queries \cite{uddin2026conformalrouting} or a cascade's per-query cost tail \cite{valkanas2025c3po}. \system instead certifies the expected normalized \emph{cost} of a pre-generation cutoff from 40 paired validation records, without a multiple-testing penalty because its audit statistic is monotone in the cutoff (Section~\ref{sec:certificate}). Output length, or at least its rank within a batch, and hence token-priced cost, is predictable from the prompt \cite{zheng2023lengthperception,jin2023s3,fu2024learningtorank}; those predictors schedule a single model, whereas \system uses the same signal to choose the model and to normalize the OSC probability by the predicted incremental price.

\section{Conclusion and Future Work}%
\label{sec:limitations}
\paragraph{Summary.}
\system shares one frozen encoder between a linear OSC head and a small cost network and chooses the answer model before generation. Fitted to each task from a small set of paired records whose cost is charged, \systemB has the highest mean accuracy at every matched cost on every task, \systemS stays within 0.6 points of its quota while exceeding the baselines at equal dollars, and fifty labels capture most of what labels can buy once their cost is charged.

\paragraph{Limitations and future work.}
Three directions follow from the results. The first is a better quality
signal: the current head predicts only whether the strong model fixes a
weak error and ignores the reverse case, and the gap to the oracle
quality signal (Table~\ref{tab:quality-signal}) shows how much accuracy remains to be gained, whether through richer query
representations or through weak supervision from unlabeled strong
outputs, of which the strong-model oracle of Section~\ref{sec:strong-label} is a first step. The second is a guarantee that reaches the deployed rule: the
certificate covers one fixed cutoff, and extending it to the paced,
adaptive rule, to dollar budgets, and to an accounting that sees only the
call actually made would close the distance between the theorem and
practice. This last gap also qualifies the quota adherence reported above: pacing charges each routed query its normalized cost (Eq.~\ref{eq:share}), which depends on the price of the call that was not made, so a cached replay can compute it but a live system cannot. The third is broader evaluation: prospective runs on fresh
workloads such as semantic filters and agent steps judged by their users rather than by reference answers, with both models' answers collected so that accuracy can still be compared at matched cost, routing among more than two models,
and learning from the training records alone when to down-weight the cost term, which lowers accuracy slightly where GPT-5's cost varies less (Table~\ref{tab:cost-awareness-main}).

\bibliographystyle{ACM-Reference-Format}
\bibliography{main}

\end{document}